\documentclass[preprintnumbers,nofootinbib,a4paper,superscriptaddress]{revtex4-1}

\usepackage{amsmath,latexsym,amssymb,amsfonts}
\usepackage{amsthm}
\usepackage{mathrsfs}
\usepackage{xcolor}

\usepackage[hidelinks]{hyperref}

\theoremstyle{definition}
\newtheorem{problem}{Problem}
\theoremstyle{plain}

\newtheorem{proposition}{Proposition}
\newcommand{\OO}{\mathcal O}
\newcommand{\PP}{\mathbb P}
\newcommand{\cL}{\mathcal L}
\allowdisplaybreaks[2]
\begin{document}

\title{Curvature Integrability in Twinor Theory\texorpdfstring{\\}{: }An Ambitwistor Approach to the Twinor Googly Problem}
\author{Chen-Hsu Chien}
\email[E-mail: ]{chien@math.muni.cz}
\affiliation{Department of Mathematics and Statistics, Masaryk University}
\date{\today}

\begin{abstract}
The twinor analogue of the googly problem is addressed through an ambitwistor structure defined by null autoparallels, retaining both chiralities, sources and torsion without requiring chiral null surfaces. Under the stated quotient, evaluation and descent assumptions, holomorphic affine sections reconstruct the twinor field equation $DY=J$. Correspondence-space cohomology realizes the contracted chiral and ladder equations, including the ladder's soldering coupling and mass-normalization source. For a prescribed non-normal conformal connection, infinity-twistor parallelism recovers the energy--momentum equation, supplemented by independent spin-current equations. These constructions give a conditional auxiliary formulation; constitutive sources and gravitational prescriptions remain dynamical input.
\end{abstract}

\maketitle

\section{Introduction}

Twinor theory describes fields through spin representations of a reductive Cartan geometry modelled on complexified $(A)dS_4$, with independent spin connection, soldering and particle scale \cite{Chien}. Its field equation $DY=J$ has curvature compatibility $DJ=F_DY$, with $F_D=D^2$ and one-form source $J$. Curvature--soldering equations specify Einstein--Cartan dynamics separately. The aim here is an auxiliary holomorphic reconstruction retaining both chiralities and torsion.

Conformal Cartan geometry provides the local twistor bundles and connections underlying the twistor description of spacetime \cite{AF,ChienConformal}. In the nonlinear-graviton construction, integrability of the chiral null-plane distribution produces curved twistor space and requires one Weyl chirality to vanish \cite{Penrose}. Ordinary ambitwistor constructions instead use null geodesics and accommodate both chiralities, with additional geometric data encoding field equations \cite{MS,LeBrun}. The present approach follows this correspondence-space perspective, replacing chiral null surfaces by rays to retain both chiralities and allow torsion.

\begin{problem}[Twinor googly problem]\label{prob:googly}
Construct an auxiliary holomorphic formulation equivalent to the twinor field equation with sources, retaining both chiralities and allowing torsion without imposing half-flatness, and specify the hypotheses under which its data reconstruct spacetime fields.
\end{problem}

The \emph{ambitwistor structure} considered here is the ray correspondence of projective null directions horizontally transported by the Lorentz spin connection. Its rays are null autoparallels; neither integrable $\alpha$- and $\beta$-surface families nor their intersections are required. Proposition~\ref{prop:reconstruction} gives a conditional resolution of Problem~\ref{prob:googly}: for fixed connection and source, normalized holomorphic affine sections reconstruct exactly the solutions of $DY=J$, assuming a simple ray quotient, complete skies and effective descent. On the correspondence space, Serre pairings realize the contracted chiral equations and extend to the coupled ladder system of Proposition~\ref{prop:ladder}. The latter combines ray differentiation with fibre differential operators, retaining the original soldering coupling and field-dependent mass-normalization source.

Infinity-twistor parallelism first selects the energy--momentum equation for the prescribed non-normal connection $p=P+Q$; spin-current equations remain independent. Generating these prescriptions from independent auxiliary data, and hence solving the coupled nonlinear gravitational system, is a stronger problem left open here.

\section{Conformal Cartan geometry and gravitational spin transport}\label{sec:geometry}

Let $M$ carry a holomorphic conformal Cartan geometry of type $(SL(4,\mathbb C),H_{\mathrm{conf}})$, where $H_{\mathrm{conf}}$ is the parabolic subgroup stabilizing the lower two-plane in $\mathbb C^4$. In a local spin frame adapted to this filtration, with a chosen splitting and scale, the Cartan connection is represented by
\begin{equation}\label{eq:conformalspin}
\mathcal A=\begin{pmatrix}\omega_L+\frac12\alpha\mathbf1&ie\\-ip&\omega_R-\frac12\alpha\mathbf1\end{pmatrix}.
\end{equation}
Here $e^{AA'}$ is nondegenerate soldering, $\omega_L,\omega_R$ are Lorentz spin connections, $\alpha$ is the dilatation one-form, and $p_{A'B}$ is the complementary conformal one-form. Neither normality nor a reality condition is imposed. The metric and contorsion are specified by
\begin{equation}\label{eq:metric}
g=\epsilon_{AB}\epsilon_{A'B'}e^{AA'}\otimes e^{BB'},\qquad
\omega=\mathring\omega+\kappa,
\end{equation}
where $\mathring\omega$ is Levi--Civita. 

The gravitational prescription fixes $p$ through the Lorentz curvature and energy--momentum current. Define
\begin{equation}\label{eq:torsion}
\begin{gathered}
R_{L,R}=d\omega_{L,R}+\omega_{L,R}\wedge\omega_{L,R},\qquad
\tau=de+\omega_L\wedge e+e\wedge\omega_R,\\
\mathcal E^{AA'}=R_L{}^A{}_B\wedge e^{BA'}+e^{AB'}\wedge R_R{}_{B'}{}^{A'}.
\end{gathered}
\end{equation}
Set $p_{A'B}=P_{A'B}+Q_{A'B}$, where
\begin{equation}\label{eq:PQ}
\begin{aligned}
P_{AB'}&=\frac{i}{2}\left[\star^{-1}\mathcal E_{AB'}-\frac13\operatorname{tr}_e(\star^{-1}\mathcal E)e_{AB'}\right],\\
Q_{AB'}&=4\pi G\left[\star^{-1}\mathcal J_{T\,AB'}-\frac13\operatorname{tr}_e(\star^{-1}\mathcal J_T)e_{AB'}\right].
\end{aligned}
\end{equation}
Here $\operatorname{tr}_eu=\operatorname{tr}(e^{-1}\circ u)$, so $\operatorname{tr}_ee=4$, and $\star^{-1}$ acts only on form degree. With $E=(\epsilon_{AB})$ and $E'=(\epsilon^{A'B'})$, the spinor-lowered soldering block is $e^\flat=-E'e^tE$, with components $e_{A'B}$. The blocks $P_{A'B},Q_{A'B}$ are obtained using the same lowering and index-order conventions. The energy--momentum current three-form is $\mathcal J_T^{AA'}=T_{BB'}{}^{AA'}\star e^{BB'}$.\footnote{The energy--momentum components are zero-forms; $\mathcal J_T$ is the Hodge dual of the soldering-valued one-form $T_{BB'}{}^{AA'}e^{BB'}$. The convention is $\star e^{AA'}=\frac{i}{3}e^{AC'}\wedge e_{C'B}\wedge e^{BA'}$.} For $\omega=\mathring\omega$, $P_{AB'}=P_{AB'CC'}e^{CC'}$ is the Schouten one-form in the curvature convention of \cite{Chien}, with $P_{AB'CC'}=P_{CC'AB'}$. This Lorentz-equivariant prescription extends locally to a conformal Cartan connection in the chosen scale and splitting; invariance under changing these choices is not asserted.

For constant $\ell\ne0$, put $\Lambda=-3/\ell^2$ (so $\Lambda\ne0$ in this reduction) and introduce the nondegenerate covariant infinity twistor
\begin{equation}\label{eq:infinitytwistor}
I_\ell=\operatorname{diag}\left(-\frac{E}{2\ell^2},E'\right).
\end{equation}
Preservation of the spin metrics gives
\begin{equation}\label{eq:spinreduction}
\mathcal DI_\ell=dI_\ell-\mathcal A^tI_\ell-I_\ell\mathcal A=0
\quad\Longleftrightarrow\quad
\alpha=0,\qquad p=-\frac{e^\flat}{2\ell^2}.
\end{equation}
With $p=P+Q$, the invertible trace adjustment in \eqref{eq:PQ} identifies the second condition with the energy--momentum equation:
\begin{equation}\label{eq:Einsteinequivalent}
P_{AB'}+Q_{AB'}=\frac{\Lambda}{6}e_{AB'}
\quad\Longleftrightarrow\quad
G(\omega)+\Lambda g=-8\pi G\,T.
\end{equation}
Indeed, for $u=i\star^{-1}\mathcal E+8\pi G\,\star^{-1}\mathcal J_T$, Eq.~\eqref{eq:PQ} gives $2(P+Q)=u-\tfrac13(\operatorname{tr}_eu)e$. Since $\operatorname{tr}_ee=4$, its inverse gives $u=2(P+Q)-2\operatorname{tr}_e(P+Q)e$, so $P+Q=\Lambda e/6$ is equivalent to $u=-\Lambda e$. Here $G(\omega)$ is the Einstein tensor of the full metric-compatible connection in the original curvature convention, including its antisymmetric part, with the paired indices ordered as in \cite[Eq.~(61)]{Chien}.

Parallelism of $I_\ell$ reduces the connection to its symplectic stabilizer. Within this stabilizer, the two nondegenerate planes in the chosen splitting have common stabilizer $H=SL(2,\mathbb C)_L\times SL(2,\mathbb C)_R$, with fundamental spin bundle $T=S^A\oplus S_{A'}$. The constant change of frame $S_\ell=\operatorname{diag}(\mathbf1,(\sqrt2\ell)^{-1}\mathbf1)$ gives
\begin{equation}\label{eq:cartan}
A=S_\ell^{-1}\mathcal A S_\ell
=\begin{pmatrix}
\omega_L&ie/(\sqrt2\ell)\\
ie^\flat/(\sqrt2\ell)&\omega_R
\end{pmatrix}.
\end{equation}
With nondegenerate soldering, this recovers the reductive twinor Cartan geometry of type $(Sp(4,\mathbb C),H)$ with its full contorsion \cite{Chien}.

The spin-current equations complete the Einstein--Cartan system:
\begin{equation}\label{eq:spincurrent}
\begin{aligned}
\tau^{AC'}\wedge e_{C'B}-e^{AC'}\wedge\tau_{C'B}
&=8\pi G\,\mathcal S_L{}^A{}_B,\\
\tau_{A'C}\wedge e^{CB'}-e_{A'C}\wedge\tau^{CB'}
&=8\pi G\,\mathcal S_R{}_{A'}{}^{B'}.
\end{aligned}
\end{equation}
For prescribed spin-current three-forms and nondegenerate soldering, these equations determine contorsion algebraically \cite{Chien,Hehl}. If the currents depend algebraically on the connection, elimination requires invertibility of the coupled algebraic system. They remain independent of infinity-twistor parallelism. In vacuum, vanishing energy--momentum and spin currents give $Q=0$, $\kappa=0$ and $p_{A'B}=P_{A'B}=\Lambda e_{A'B}/6$, recovering the normal conformal spin connection of an Einstein metric without eliminating either Weyl chirality.

\section{The ambitwistor structure}

The Cartan geometry of Section~\ref{sec:geometry} determines the bundle of projective null directions
\begin{equation}\label{eq:incidence}
\pi:\Sigma=\PP(S)\times_M\PP(S')\longrightarrow M,\qquad
\mathcal C_x=\pi^{-1}(x)\simeq\mathbb{CP}^1\times\mathbb{CP}^1.
\end{equation}
At $z=(x,[\xi],[\eta])\in\Sigma$, let $k=\xi^A\eta^{A'}e_{AA'}$, where $e_{AA'}$ denotes the dual frame. The spin connection $\omega$ defines the unique horizontal lift $k^{\mathrm{hor}}$ satisfying $d\pi_z(k^{\mathrm{hor}})=k$, and hence the rank-one distribution $\cL_z=\mathbb Ck^{\mathrm{hor}}\subset T_z\Sigma$. Integral curves admit local representatives and a parameter satisfying
\begin{equation}\label{eq:raytransport}
\dot x^{AA'}=\xi^A\eta^{A'},\qquad
\nabla^\omega_{\dot x}\xi^A=0,\qquad
\nabla^\omega_{\dot x}\eta^{A'}=0
\quad\Longrightarrow\quad
\nabla^\omega_{\dot x}\dot x=0.
\end{equation}
Their projections are null autoparallels. Assume a holomorphic submersion onto a Hausdorff ray space whose connected fibres are the leaves of $\cL$,
\begin{equation}\label{eq:rayquotient}
M\xleftarrow{\ \pi\ }\Sigma\xrightarrow{\ q\ }\mathcal T,\qquad
\ker dq=\cL,\qquad \dim_{\mathbb C}\mathcal T=5.
\end{equation}
Each point of $\mathcal T$ represents an unparametrized ray. With vanishing torsion, this is the ordinary complex null-geodesic space. More generally, $\mathring\nabla_kk=-\kappa(k)k$ along an affinely parametrized autoparallel, so the rays coincide with metric null geodesics as unparametrized curves precisely when $\kappa(k)k\in\mathbb Ck$ for every null $k$. A canonical contact structure is not assumed for the torsionful ray space.

Require $q$ to embed each complete quadric $\mathcal C_x$. Its image is the \emph{sky} of $x$, the family of rays through that point \cite{NatarioTod}. With $\operatorname{pr}_L,\operatorname{pr}_R$ the projections onto the two factors,
\begin{equation}\label{eq:skyweight}
\begin{gathered}
\cL|_{\mathcal C_x}\simeq\OO(-1,-1)
=\operatorname{pr}_L^*\OO(-1)\otimes\operatorname{pr}_R^*\OO(-1),\\
\OO(-1,-1)_{([\xi],[\eta])}=\mathbb C\xi\otimes\mathbb C\eta,\qquad
H^0(\mathcal C_x,\cL^*)\simeq T_x^*M.
\end{gathered}
\end{equation}
The first identification follows from soldering and horizontal lifting; the last sends $a_{AA'}$ to the section $a_{AA'}\xi^A\eta^{A'}$ of $\OO(1,1)$.

This correspondence with its induced ray transport is called the \emph{ambitwistor structure}. Rank-one involutivity imposes no Weyl-chirality restriction; field equations require compatibility across the ray family. The quotient with complete embedded skies is a separate hypothesis, stronger than local flow-box existence.

For a bundle $E$ with connection, assume effective holomorphic descent of ray transport and holomorphically varying sky evaluation:
\begin{equation}\label{eq:rayevaluation}
q^*E_{\mathcal T}\simeq\pi^*E,\qquad
E_x\simeq H^0(\mathcal C_x,E_{\mathcal T}|_{\mathcal C_x}).
\end{equation}
The first identification is the evaluation isomorphism and respects transport; restriction means pullback along $q|_{\mathcal C_x}$. A holomorphic section of $E_{\mathcal T}$ is constant on each complete sky in the evaluation trivialization and reconstructs a spacetime section parallel in every null direction, hence fully parallel. Conversely, a parallel section descends under these assumptions.

Applied to the alternating-form bundle of \eqref{eq:conformalspin}, this identifies a holomorphic section evaluating to $I_\ell$ with $\mathcal DI_\ell=0$. The prescription \eqref{eq:PQ} then selects the energy--momentum equation, while \eqref{eq:spincurrent} remains additional input. Gravitational reconstruction therefore uses the prescribed transport data together with the ray correspondence.

\section{The twinor field equation and affine reconstruction}

For $2s\in\mathbb N_{>0}$, let $E_s=\operatorname{Sym}^{2s}T\otimes V$ carry the coupled twinor connection
\begin{equation}\label{eq:connections}
D=d+\rho_{2s*}(A_\lambda)\otimes\mathbf1_V
+\mathbf1_{\operatorname{Sym}^{2s}T}\otimes a_V,\qquad
A_\lambda=
\begin{pmatrix}
\omega_L&ie/(2\sqrt2\lambda)\\
ie^\flat/(2\sqrt2\lambda)&\omega_R
\end{pmatrix}.
\end{equation}
Here $\rho_{2s*}$ is the induced action on the symmetric twinor representation, $a_V$ is the internal connection, and $\lambda\ne0$ is the particle scale, independent of $\ell$. The fundamental Dirac twinor corresponds to $2s=1$, and higher symmetric powers give the ladder representations. The full connection $\omega=\mathring\omega+\kappa$ is retained.

For a holomorphic field $Y\in\Omega^0(M,E_s)$ and one-form source $J\in\Omega^1(M,E_s)$, the twinor field equation and its necessary curvature compatibility are
\begin{equation}\label{eq:field}
DY=J,\qquad DJ=F_DY,\qquad F_D=D^2,
\end{equation}
where $\Omega^r(M,E_s)$ denotes holomorphic $E_s$-valued $r$-forms and $D$ also denotes the exterior covariant derivative. Constitutive relations for $J$ supply the matter dynamics: unrestricted assignment of $J=DY$ imposes no equation on $Y$. Curvature compatibility requires neither flatness nor a homogeneous Buchdahl condition on the unrestricted field--source pair. Prescribed source relations may nevertheless impose further curvature restrictions \cite[Eqs.~(69)--(72)]{Chien}.

The field equation is equivalent to parallel transport of $(Y,1)$ under the affine connection
\begin{equation}\label{eq:affine}
\widehat D=\begin{pmatrix}D&-J\\0&d\end{pmatrix},\qquad
\widehat D\binom Y1=0,\qquad
\widehat F=\begin{pmatrix}F_D&-DJ\\0&0\end{pmatrix}.
\end{equation}
Thus $\widehat F(Y,1)^t=0$ gives the second relation in \eqref{eq:field}; this necessary condition alone does not guarantee solvability for prescribed $J$.

On the ray space $\mathcal T$, consider the holomorphic vector-bundle extension
\begin{equation}\label{eq:extension}
0\longrightarrow E_{s,\mathcal T}
\xrightarrow{\iota}\widehat E_{s,\mathcal T}
\xrightarrow{\operatorname{pr}}\OO_{\mathcal T}
\longrightarrow0.
\end{equation}
In a local splitting, $\iota(y)=(y,0)$ and $\operatorname{pr}(y,c)=c$, with $\ker\operatorname{pr}=\operatorname{im}\iota$. The enlarged bundle encodes the auxiliary scalar component of $(Y,1)$; the fibre $\operatorname{pr}^{-1}(1)$ is affine, modelled on $E_{s,\mathcal T}$.

Assume effective descent, sky triviality and holomorphic evaluation for the entire extension, with its maps respecting ray transport. The difference between its partial connection and a reference connection lies in
\begin{equation*}
H^0\!\left(\mathcal C_x,\operatorname{End}(\widehat E_{s,x})\otimes\cL^*\right)
\simeq\operatorname{End}(\widehat E_{s,x})\otimes T_x^*M,
\end{equation*}
so \eqref{eq:skyweight} reconstructs a unique full connection. Require its subbundle and quotient connections to be $D$ and $d$. Over Stein $U\subset M$, Cartan's theorem B \cite{CartanB} gives $H^1(U,\OO(E_s))=0$, hence a holomorphic splitting $\widehat E_s|_U\simeq E_s|_U\oplus\OO_U$. In the prescribed physical splitting the remaining connection column is required to equal $-J$, as in \eqref{eq:affine}.

\begin{proposition}[Conditional twinor reconstruction] \label{prop:reconstruction}
For fixed holomorphic $(D,J)$ and the stated transport-compatible descent, evaluation and splitting, normalized sections
\begin{equation*}
\widehat{\sigma}\in H^0(\mathcal T,\widehat E_{s,\mathcal T}),
\qquad \operatorname{pr}(\widehat{\sigma})=1,
\end{equation*}
correspond bijectively to holomorphic solutions of $DY=J$. Such sections exist precisely when the extension class of \eqref{eq:extension} vanishes in $H^1(\mathcal T,\OO(E_{s,\mathcal T}))$.
\end{proposition}

\begin{proof}
Sky evaluation reconstructs $(Y,1)$ from $q^*\widehat{\sigma}$, which is constant on each compact connected sky in the evaluation trivialization. Relative parallelism gives $(DY-J)(k)=0$ for every null $k$, hence $DY=J$. Conversely, affine parallel fields descend, and the constructions are inverse. A normalized section splits \eqref{eq:extension} by $c\mapsto c\,\widehat{\sigma}$, proving the extension-class criterion.
\end{proof}

\section{Chiral fields and the Penrose construction with sources}

The Penrose contour representation \cite{PenroseZRM,EPW} motivates a cohomological realization of the contracted chiral twinor equations on the correspondence space $\Sigma$. The fields $W^{(2s,0)}$ and $W^{(0,2s)}$ take values in $\operatorname{Sym}^{2s}S\otimes V$ and $\operatorname{Sym}^{2s}S'^*\otimes V$. Their derivative $D$ contains the Lorentz and internal connections, including contorsion. The particle soldering coupling is absent because it mixes Lorentz components and preserves neither chiral representation; soldering still identifies tangent and spinor indices and defines the ray distribution.

Let $\mathfrak l_L,\mathfrak l_R$ be the tautological spinor lines on $\Sigma$. For $2s\in\mathbb N_{>0}$, define
\begin{equation}\label{eq:weights}
\begin{aligned}
V_L&=\mathfrak l_L^{\otimes(2s+2)},&
V_L|_{\mathcal C_x}&=\OO(-2s-2,0),&
(V_L\otimes\cL^*)|_{\mathcal C_x}&=\OO(-2s-1,1),\\
V_R&=\mathfrak l_R^{\otimes(2s+2)},&
V_R|_{\mathcal C_x}&=\OO(0,-2s-2),&
(V_R\otimes\cL^*)|_{\mathcal C_x}&=\OO(1,-2s-1).
\end{aligned}
\end{equation}
Tensoring with $\pi^*V$ and its induced transport supplies the internal coefficients, suppressed below. Serre duality \cite[Ch.~VI, Sec.~7]{Demailly}, with canonical bundle $\OO(-2)$ on $\mathbb{CP}^1$, identifies $H^1(\OO(-2s-2))$ with the dual of the degree-$2s$ homogeneous polynomials. Consequently,
\begin{equation}\label{eq:chiralcohomology}
H^1(\mathcal C_x,V_L)\simeq\operatorname{Sym}^{2s}S_x,\qquad
H^1(\mathcal C_x,V_L\otimes\cL^*)\simeq
\operatorname{Sym}^{2s-1}S_x\otimes S_x'^*,
\end{equation}
with the primed identifications obtained by exchanging the rulings and applying the spin metrics.

Over a Stein region $U\subset M$, properness of $\pi$ and concentration of the fibre cohomology in degree one identify correspondence-space classes with holomorphic field and contracted-source sections by the Leray spectral sequence \cite[Ch.~IV, Sec.~13]{Demailly}. For representatives $f,b$, the cohomological equation is
\begin{equation}\label{eq:cochains}
[\nabla_{\cL}f-b]=0
\quad\Longleftrightarrow\quad
\nabla_{\cL}f=b+\delta h.
\end{equation}
Here $\nabla_{\cL}$ is the induced partial covariant derivative, a $\mathbb C$-linear sheaf differential operator, not an $\OO_\Sigma$-linear map; $f,b$ are degree-one \v{C}ech cocycles, and $h$ is a degree-zero cochain on a common acyclic cover, refined if necessary. Appendix~\ref{app:cohomology} explains these representatives and their reconstruction.

The fibrewise Serre pairing reconstructs
\begin{equation}\label{eq:penrose}
W^{(2s,0)\,A_1\cdots A_{2s}}(x)
=\frac{1}{2\pi i}\oint
\xi^{A_1}\cdots\xi^{A_{2s}}f_x(\xi,\eta)\,\xi_Bd\xi^B,
\end{equation}
where a representative independent of the second ruling may be chosen on each sky. Coboundaries have zero pairing, so
\begin{equation}\label{eq:sourcepairing}
\begin{aligned}
j_{L\,B'}{}^{A_2\cdots A_{2s}}\eta^{B'}
&=\frac{1}{2\pi i}\oint
\xi^{A_2}\cdots\xi^{A_{2s}}b_x\,\xi_Cd\xi^C\\
&=\frac{1}{2\pi i}\oint
\xi^{A_2}\cdots\xi^{A_{2s}}\xi^B\eta^{B'}D_{BB'}f_x\,\xi_Cd\xi^C\\
&=\eta^{B'}D_{BB'}W^{(2s,0)\,BA_2\cdots A_{2s}}.
\end{aligned}
\end{equation}
The middle expression is evaluated in spin and internal frames with connection coefficients zero at $x$, differentiating a representative over a neighbourhood before restriction to the sky. No coframe derivative enters; covariance restores the full torsionful connection.

Together with the primed pairing, this proves
\begin{equation}\label{eq:chiral}
\begin{aligned}
{}[\nabla_{\cL}f_L-b_L]=0
&\quad\Longleftrightarrow\quad
D_{A_1B'}W^{(2s,0)\,A_1\cdots A_{2s}}
=j_{L\,B'}{}^{A_2\cdots A_{2s}},\\
[\nabla_{\cL}f_R-b_R]=0
&\quad\Longleftrightarrow\quad
D^{BA'_1}W^{(0,2s)}{}_{A'_1\cdots A'_{2s}}
=j_R{}^B{}_{A'_2\cdots A'_{2s}}.
\end{aligned}
\end{equation}
The converse follows from the same cohomological identifications: equality of the reconstructed sections implies equality of their classes. These are contracted equations; agreement with a prescribed full one-form source additionally requires its complementary symmetric projection. Constitutive source relations supply the dynamics. The construction is on $\Sigma$; an interpretation as cohomology on the ray space $\mathcal T$ requires additional descent and comparison results.

\section{Ladder fields and coupled cohomological reconstruction}

The ladder bundle $\operatorname{Sym}^{2s}T\otimes V$ has components $Y^{(2s-k,k)}{}_{A'_1\cdots A'_k}{}^{A_{k+1}\cdots A_{2s}}$. Define
\begin{equation}\label{eq:ladderweights}
\mathscr V_k
=\mathfrak l_L^{\otimes(2s-k+2)}
\otimes(\mathfrak l_R^*)^{\otimes k}
\otimes\pi^*V,\qquad
0\leq k\leq 2s.
\end{equation}
Suppressing the internal factor, $\mathscr V_k|_{\mathcal C_x}=\OO(-2s+k-2,k)$, and Serre duality on the first ruling gives
\begin{equation}\label{eq:laddercohomology}
H^1(\mathcal C_x,\mathscr V_k)
\simeq
\operatorname{Sym}^{2s-k}S_x
\otimes\operatorname{Sym}^{k}S_x'^*\otimes V_x.
\end{equation}
Every ladder component is therefore represented in degree one. The choice of ruling fixes the cohomological representation, not a chirality restriction.

A representative $f_k$, homogeneous of degrees $(-2s+k-2,k)$ in $(\xi,\eta)$, reconstructs
\begin{equation}\label{eq:ladderpairing}
Y^{(2s-k,k)}{}_{A'_1\cdots A'_k}{}^{A_{k+1}\cdots A_{2s}}(x)
\,\eta^{A'_1}\cdots\eta^{A'_k}
=
\frac{1}{2\pi i}\oint
\xi^{A_{k+1}}\cdots\xi^{A_{2s}}f_{k,x}(\xi,\eta)\,\xi_Bd\xi^B.
\end{equation}
The dependence on $\eta$ supplies the primed indices.\footnote{The primed contour pairing uses a new representative $\widetilde f_{k,x}$ following:
\begin{equation*}
Y^{(p,k)}{}_{A'_1\cdots A'_k}{}^{A_{k+1}\cdots A_{2s}}(x)\,\xi_{A_{k+1}}\cdots\xi_{A_{2s}}
=
\frac{1}{2\pi i}\oint_{\Gamma_\eta}\eta_{A'_1}\cdots\eta_{A'_k}\,\widetilde f_{k,x}(\xi,\eta)\,\eta_{B'}d\eta^{B'}.
\end{equation*}} Over Stein $U\subset M$, these pairings identify $H^1(\pi^{-1}(U),\OO(\mathscr V_k))$ with holomorphic ladder-component sections.

Let $\nabla_{BB'}$ denote the Lorentz and internal derivative, acting horizontally on homogeneous representatives; the particle soldering coupling is included separately below. In a local spin and internal frame, its action is
\begin{equation}\label{eq:horizontalrepresentative}
\nabla_{BB'}f=e_{BB'}f
-\omega_{L,BB'}{}^A{}_C\xi^C\frac{\partial f}{\partial\xi^A}
-\omega_{R,BB'}{}^{A'}{}_{C'}\eta^{C'}\frac{\partial f}{\partial\eta^{A'}}
+a_{V,BB'}f,
\end{equation}
where $e_{BB'}$ differentiates the base variables at fixed fibre coordinates and $\omega_R{}^{A'}{}_{C'}$ acts on the raised primed spinors. The pair $BB'$ labels the coframe component $\omega_{L,BB'}{}^A{}_C:=\omega_L{}^A{}_C(e_{BB'})$ in $\omega_L{}^A{}_C=e^{BB'}\omega_{L,BB'}{}^A{}_C$, and the same convention applies to $\omega_R$ and $a_V$ throughout. All subsequent fibre derivatives act after this horizontal derivative. Define
\begin{equation}\label{eq:ladderoperators}
\begin{aligned}
\mathscr L_k f
&=\xi^B\eta^{B'}\nabla_{BB'}f,
&&0\leq k<2s,\\
\mathscr R_k f
&=-\frac{1}{(2s-k+1)k}
\frac{\partial}{\partial\xi^B}
\frac{\partial}{\partial\eta^{B'}}
\bigl(\nabla^{BB'}f\bigr),
&&1\leq k\leq 2s,\\
\mathscr M_k f
&=\frac1k\,\xi^B\frac{\partial}{\partial\eta^{B'}}
\bigl(\nabla_B{}^{B'}f\bigr),
&&1\leq k<2s.
\end{aligned}
\end{equation}
Fibre differentiation is taken on $(S\setminus0)\times_M(S^{\prime}\setminus0)$. Under spin-frame changes, the vertical derivatives transform dually to $(\xi,\eta)$, while \eqref{eq:horizontalrepresentative} absorbs the derivatives of the frame changes. The contracted expressions therefore define $\mathbb C$-linear sheaf differential operators of the stated weights and commute with the \v{C}ech differential. Their target weights are respectively $(-2s+k-1,k+1)$, $(-2s+k-3,k-1)$ and $(-2s+k-1,k-1)$. In particular, $\mathscr L_k=\nabla_{\cL}$ maps $\mathscr V_k$ to $\mathscr V_{k+1}$, while $\mathscr R_k$ maps $\mathscr V_k$ to $\mathscr V_{k-1}$.

Under \eqref{eq:ladderpairing}, these operators induce the original adjacent and mixed contractions:
\begin{equation}\label{eq:ladderprojections}
\begin{aligned}
{}[\mathscr L_kf_k]&\longleftrightarrow
\operatorname{ctr}_L\bigl(\nabla Y^{(2s-k,k)}\bigr),\\
[\mathscr R_kf_k]&\longleftrightarrow
\operatorname{ctr}_R\bigl(\nabla Y^{(2s-k,k)}\bigr),\\
[\mathscr M_kf_k]&\longleftrightarrow
\operatorname{ctr}_0\bigl(\nabla Y^{(2s-k,k)}\bigr).
\end{aligned}
\end{equation}
The contractions and spinor lowering are those of \cite[Eqs.~(27)--(29)]{Chien}. The forward identity follows as in \eqref{eq:sourcepairing}. For the reverse identity, integration by parts gives
\begin{equation}\label{eq:ladderintegration}
\frac{1}{2\pi i}\oint\xi^{A_k}\cdots\xi^{A_{2s}}
\frac{\partial f_k}{\partial\xi^B}\,\xi_Cd\xi^C=-(2s-k+1)\delta_B^{(A_k}
\frac{1}{2\pi i}\oint\xi^{A_{k+1}}\cdots\xi^{A_{2s})}f_k\,\xi_Cd\xi^C,
\end{equation}
while differentiation in $\eta$ contributes the factor $k$. The mixed identity combines multiplication by $\xi$ with differentiation in $\eta$.

The full particle connection \eqref{eq:connections} contributes the adjacent terms
\begin{equation}\label{eq:laddersoldering}
\begin{aligned}
\operatorname{ctr}_L\bigl((DY)^{(2s-k,k)}\bigr)
&=\operatorname{ctr}_L\bigl(\nabla Y^{(2s-k,k)}\bigr)
+\frac{i(2s-k+1)}{2\sqrt2\lambda}Y^{(2s-k-1,k+1)},\\
\operatorname{ctr}_R\bigl((DY)^{(2s-k,k)}\bigr)
&=\operatorname{ctr}_R\bigl(\nabla Y^{(2s-k,k)}\bigr)
+\frac{i(k+1)}{2\sqrt2\lambda}Y^{(2s-k+1,k-1)}.
\end{aligned}
\end{equation}
The prescribed mass-normalization source is \cite[Eq.~(37)]{Chien}
\begin{equation}\label{eq:laddermasssource}
\begin{aligned}
(j_{\rm mass}[Y])_{L,k}&=\frac{i(2s-k-1)}{2\sqrt2\lambda}Y^{(2s-k-1,k+1)},&&0\leq k<2s,\\
(j_{\rm mass}[Y])_{R,k}&=\frac{i(k-1)}{2\sqrt2\lambda}Y^{(2s-k+1,k-1)},&&1\leq k\leq2s.
\end{aligned}
\end{equation}
Subtracting these field-dependent terms leaves $i/(\sqrt2\lambda)$ in both adjacent equations; the soldering contribution to the mixed projection is zero. Consequently, the coupled cohomological system is
\begin{equation}\label{eq:laddercech}
\begin{aligned}
\left[\mathscr L_kf_k+\frac{i}{\sqrt2\lambda}f_{k+1}-b_{L,k}\right]&=0,
&&0\leq k<2s,\\
\left[\mathscr R_kf_k+\frac{i}{\sqrt2\lambda}f_{k-1}-b_{R,k}\right]&=0,
&&1\leq k\leq 2s,\\
[\mathscr M_kf_k-b_{0,k}]&=0,
&&1\leq k<2s.
\end{aligned}
\end{equation}
The source cocycles $b_{L,k},b_{R,k},b_{0,k}$ take values in $\mathscr V_{k+1}$, $\mathscr V_{k-1}$ and $\mathscr V_k\otimes\mathfrak l_L^*\otimes\mathfrak l_R$, respectively, and reconstruct the physical contracted sources $j_{L,k},j_{R,k},j_{0,k}$. Their classes are prescribed; cocycle representatives are immaterial. Each equality means that its representative is a \v{C}ech coboundary, as in \eqref{eq:cochains}.

\begin{proposition}[Contracted ladder correspondence]\label{prop:ladder}
Over a Stein spacetime region, the pairings \eqref{eq:ladderpairing} identify cohomology classes $([f_k])_{k=0}^{2s}$ satisfying \eqref{eq:laddercech} with holomorphic solutions of
\begin{equation}\label{eq:ladderfield}
\begin{aligned}
\operatorname{ctr}_L\bigl(\nabla Y^{(2s-k,k)}\bigr)
+\frac{i}{\sqrt2\lambda}Y^{(2s-k-1,k+1)}
&=j_{L,k},
&&0\leq k<2s,\\
\operatorname{ctr}_R\bigl(\nabla Y^{(2s-k,k)}\bigr)
+\frac{i}{\sqrt2\lambda}Y^{(2s-k+1,k-1)}
&=j_{R,k},
&&1\leq k\leq 2s,\\
\operatorname{ctr}_0\bigl(\nabla Y^{(2s-k,k)}\bigr)
&=j_{0,k},
&&1\leq k<2s.
\end{aligned}
\end{equation}
The connection, soldering and source data are fixed, with the original field-dependent mass-normalization law.
\end{proposition}

\begin{proof}
The Serre and Leray identifications reconstruct each ladder component and contracted source uniquely from its class. Equation~\eqref{eq:ladderprojections} identifies the induced differential operators, and \eqref{eq:laddersoldering} supplies the algebraic couplings. Equality of reconstructed sections is therefore equivalent to equality of the corresponding classes.
\end{proof}

The construction retains both chiralities and arbitrary holomorphic contorsion. The reverse and mixed operators use the fibre geometry of $\Sigma$ in addition to ray transport; rank-one involutivity alone does not establish the coupled equations. Recovering a prescribed full equation $DY=J$ also requires agreement of the symmetric source projection, and its curvature compatibility remains $DJ=F_DY$. Proposition~\ref{prop:reconstruction} supplies the corresponding conditional affine formulation on $\mathcal T$; descent of the present coupled cohomological operators is not asserted. Neither construction requires ladder fields to factor into fundamental spacetime solutions. The coupled cohomological equations are not asserted to form a differential complex with square zero; their differential consequences retain the curvature and source compatibility.

\section{Discussion and conclusion}

The ambitwistor structure provides a conditional resolution of the twinor reconstruction problem in Problem~\ref{prob:googly}. Its null autoparallels retain both chiralities and possible torsion without requiring integrable chiral null surfaces. Under the quotient, evaluation and descent hypotheses, normalized holomorphic affine sections reconstruct the full equation $DY=J$, with curvature compatibility $DJ=F_DY$.

The correspondence-space constructions identify the contracted chiral and ladder equations with coupled \v{C}ech relations. For the ladder, fibre differential operators complement ray transport and reproduce the adjacent soldering couplings, mass normalization and mixed contraction. These equivalences hold over Stein spacetime regions; they do not require descent to the ray space. Recovering a prescribed full source additionally requires its symmetric projection, and descent of the coupled cohomological operators remains open. In the gravitational sector, in the chosen splitting, infinity-twistor parallelism fixes $\alpha=0$ and, for $p=P+Q$, imposes the energy--momentum equation, while independent spin-current equations determine contorsion under the stated algebraic hypotheses.

Ordinary ambitwistor geometry already accommodates both chiralities \cite{MS,LeBrun}. The contribution here is its adaptation to twinor transport and coupled field representations with sources, independent particle and cosmological scales, and possible torsion. The ray-space formulation requires complete skies and effective descent; rank-one involutivity alone guarantees neither these hypotheses nor dynamical compatibility. Torsion can alter the null autoparallels of the connection. Even when their paths are unchanged, parallel transport along them can differ. Deriving the constitutive sources and gravitational prescriptions from independent auxiliary data remains a stronger problem beyond the reconstruction established here.

\appendix
\section{Cohomology and field reconstruction}\label{app:cohomology}

The cohomology used here describes holomorphic representatives modulo changes that leave their reconstructed fields unchanged. For a holomorphic bundle $E$ on $X$, let $\OO(E)$ denote its sheaf of holomorphic sections. On an acyclic cover $\{U_i\}$, a degree-zero \v{C}ech cochain is a collection of sections $a_i$ on patches, while a degree-one cochain is a collection $f_{ij}$ on overlaps. The differential is
\begin{equation}\label{eq:app_cech}
(\delta a)_{ij}=a_j-a_i,\qquad
(\delta f)_{ijk}=f_{jk}-f_{ik}+f_{ij},\qquad
\delta^2=0.
\end{equation}
Restrictions to common intersections and bundle transition functions are understood. Thus
\begin{equation}\label{eq:app_cohomology}
H^1(X,\OO(E))
=
\frac{\ker(\delta:C^1\to C^2)}
{\operatorname{im}(\delta:C^0\to C^1)}.
\end{equation}
A cocycle satisfies $\delta f=0$, and $f$ and $f+\delta a$ represent the same class. The degree counts overlaps, not spacetime differential-form degree.

For the ladder construction, take $X=\pi^{-1}(U)$ and $E=\mathscr V_k$ from \eqref{eq:ladderweights}. The sky cohomology is
\begin{equation}\label{eq:app_fibrecohomology}
\begin{aligned}
H^1(\mathcal C_x,\mathscr V_k|_{\mathcal C_x})
&\simeq
H^1(\mathbb{CP}^1,\OO(-2s+k-2))
\otimes H^0(\mathbb{CP}^1,\OO(k))\otimes V_x\\
&\simeq
\operatorname{Sym}^{2s-k}S_x
\otimes\operatorname{Sym}^{k}S_x'^*\otimes V_x.
\end{aligned}
\end{equation}
A class on one sky therefore specifies the value of $Y^{(2s-k,k)}{}_{A'_1\cdots A'_k}{}^{A_{k+1}\cdots A_{2s}}$ at $x$. Over the Stein region used in the construction, the Leray identification assembles these values into a holomorphic field:
\begin{equation}\label{eq:app_leray}
H^1(\pi^{-1}(U),\OO(\mathscr V_k))
\simeq
\Gamma_{\mathrm{hol}}\!\left(
U,\operatorname{Sym}^{2s-k}S
\otimes\operatorname{Sym}^{k}S'^*\otimes V
\right).
\end{equation}
This identifies field data; the differential equations are imposed separately.

Concretely, in local spin and internal frames, trivializing the first weight by $\xi^0=1$ and writing $z=\xi^1/\xi^0$, a fibre class admits a representative whose nontrivial Laurent part is
\begin{equation}\label{eq:app_laurent}
f_{k,x}(z,\eta)
=
\sum_{r=0}^{2s-k}
\frac{C_{r\,A'_1\cdots A'_k}(x)\eta^{A'_1}\cdots\eta^{A'_k}}{z^{r+1}}.
\end{equation}
Terms removable by sections on either projective chart are coboundaries. The remaining Laurent coefficients supply the symmetric unprimed components, while the degree-$k$ polynomials in $\eta$ supply the primed components. The Serre pairing \eqref{eq:ladderpairing} extracts these coefficients and vanishes on coboundaries. The chiral construction uses the same pairing with the weights \eqref{eq:weights}.

The equation $[\nabla_{\cL}f-b]=0$ is an equality in the source cohomology group. Since $\nabla_{\cL}$ respects restrictions, it commutes with $\delta$; hence $\delta f=\delta b=0$ implies that $\nabla_{\cL}f-b$ is a cocycle. Its class vanishes precisely when
\begin{equation}\label{eq:app_residual}
\nabla_{\cL}f_{ij}-b_{ij}=h_j-h_i
\end{equation}
for patchwise source-weight sections $h_i$. Applying the Serre pairing removes the right-hand side and gives the contracted chiral equation \eqref{eq:chiral}. Conversely, the cohomological identification implies that a vanishing reconstructed residual has zero class. The ladder equations \eqref{eq:laddercech} have the same interpretation; for example,
\begin{equation}\label{eq:app_ladderresidual}
\mathscr L_kf_{k,ij}
+\frac{i}{\sqrt2\lambda}f_{k+1,ij}
-b_{L,k,ij}
=
h_{L,k,j}-h_{L,k,i}.
\end{equation}
Thus the field class need not vanish: the class of its equation's residual must vanish. The cochains $h$ express this equality and introduce no additional spacetime fields.

The identity $\delta^2=0$ concerns overlap consistency and does not require the physical connection to be flat. Its curvature remains $D^2=F_D$, with compatibility $DJ=F_DY$. A distinct use of cohomology occurs in \eqref{eq:extension}: its extension class measures the obstruction to a global normalized affine section. Under Proposition~\ref{prop:reconstruction}, vanishing of that obstruction is equivalent to existence of a solution of the prescribed full twinor field equation.
\begin{acknowledgments}

I thank Maciej Dunajski for discussions during the CaLIGOLA Dissemination and Conclusion Workshop in Valtice, from which the question of a twinor analogue of the googly problem emerged.

I was supported by the Czech Science Foundation (GA\v{C}R) under the research project ``Cartan supergeometries and Higher Cartan geometries'' (Grant No.~GA24-10887S) and by the European Union's Horizon Europe research and innovation programme under the Marie Sk\l{}odowska-Curie grant agreement No.~101086123 (CaLIGOLA). This research was supported by the Simons Foundation grant SFI-MPS-T-Institutes-00010825 and by State Treasury funds as part of a task commissioned by the Minister of Science and Higher Education under the project ``Organization of the Simons Semesters at the Banach Center --- New Energies in 2026--2028'' (MNiSW/2025/DAP/491).

\end{acknowledgments}

\end{document}